\documentclass{svproc}
\usepackage{url}
\usepackage{graphicx}
\usepackage{enumerate}

\begin{document}
\mainmatter              
\title{Toward porting Astrophysics Visual Analytics Services to the European Open Science Cloud}
\titlerunning{Toward porting Astro Visual Analytics Services to EOSC}  
%
\author{Eva Sciacca\inst{1} \and Fabio Vitello\inst{1} \and Ugo Becciani\inst{1} \and Cristobal Bordiu\inst{1} \and Filomena Bufano\inst{1} \and
Antonio Calanducci\inst{1} \and Alessandro Costa\inst{1} \and Mario Raciti\inst{2} \and  Simone Riggi\inst{1}}
\authorrunning{Eva Sciacca et al.} 
%
\tocauthor{Eva Sciacca, Fabio Vitello, Ugo Becciani, Filomena Bufano, Antonio Calanducci, Alessandro Costa, Simone Riggi}
%

\institute{INAF, Catania Astrophysical Observatory, Catania, Italy\\
\email{eva.sciacca@inaf.it}\\
WWW home page: \texttt{https://www.researchgate.net/profile/Eva\_Sciacca}\and
Università di Catania, Dipartimento di Matematica e Informatica, Catania, Italy
}

\maketitle

\begin{abstract}
The European Open Science Cloud (EOSC) aims to create a federated environment for hosting and processing research data to support science in all disciplines without geographical boundaries, such that data, software, methods and
publications can be shared as part of an Open Science community of practice. This work presents the ongoing activities related to the implementation of visual analytics services, integrated into EOSC, towards addressing the diverse astrophysics user communities needs. These services rely on visualisation to manage the data life cycle process under FAIR principles, integrating data processing for imaging and multidimensional map creation and mosaicing, and applying machine learning techniques for detection of structures in large scale multidimensional maps.
\end{abstract}
\section{Introduction}
The European Open Science Cloud\footnote{EOSC web page: \url{https://ec.europa.eu/research/openscience/index.cfm?pg=open-science-cloud}} (EOSC) initiative has been proposed by the European Commission in 2016 to build a competitive data and knowledge economy in Europe with the vision of enabling a new paradigm of transparent, data-driven science as well as accelerating innovation driven by Open Science~\cite{ayris2016realising}.

In Astrophysics, data (and metadata) management, mapping and structure detection are fundamental tasks involving several scientific and technological challenges. A typical astrophysical data infrastructure includes several components: very large observatory archives and surveys, rich databases containing several types of metadata (e.g. describing a multitude of observations) frequently produced through long and complex pipelines, linking to outcomes within scientific publications as well as journals and bibliographic databases. In this context, visualisation plays a fundamental role throughout the data life-cycle in astronomy and astrophysics, starting from research planning, and moving to observing processes or simulation runs, quality control, qualitative knowledge discovery and quantitative analysis. The main challenges came to integrate visualisation services within common scientific workflows in order to provide appropriate supporting mechanisms for data findability, accessibility, interoperability and reusability (FAIR principles \cite{wilkinson2016fair}). 

Large-scale sky surveys are usually composed of large numbers of individual tiles --2D images or 3D data cubes--, each one mapping a limited portion of the sky. This tessellation derives from the observing process itself, when a telescope with a defined field of view is used to map a wide region of the sky by performing multiple pointings. Although it is simpler for an astronomer to handle single-pointing datasets for analysis purposes, it strongly limits the results for objects extending over multiple adjacent tiles and hampers the possibility to have a large-scale view on a particular phenomenon (e.g. the Galactic diffuse emission). Tailored services are required to map and mosaic such data for scientific exploitation in a way that their native characteristics (both in 2D and 3D) are preserved. 

Additionally, the astrophysics community produces data at very high rates, and the quantity of collected and stored data is increasing at a much faster rate than the ability to analyse them in order to find specific structures to study. Due to the sharp increase on data volume and complexity, a suite of automated structure detection services exploiting machine learning techniques is required -- consider as an example the ability to recover and classify diffuse emission and to extract compact and extended sources. 

This work presents the ongoing activities related to the implementation of services, integrated into EOSC, towards addressing the diverse astrophysics user communities needs for: 

\begin{enumerate}[(i)]
    \item putting visualisation at the centre of the data life cycle process while underpinning this by FAIR principles.
    \item integrating data processing for imaging and multidimensional map creation and mosaicing.
    \item exploiting machine learning techniques for automated structure detection in large scale multidimensional maps.
\end{enumerate}

\section{Background and Related Works}

Innovative developments in data processing, archiving, analysis and visualisation are nowadays unavoidable to deal with the data deluge expected in next-generation facilities for astronomy, such as the Square Kilometer Array\footnote{SKA web page: \url{https://www.skatelescope.org/}} (SKA). 

The increased size and complexity of the archived image products will raise significant challenges in the source extraction and cataloguing stage, requiring more advanced algorithms to extract scientific information in a mostly automated way. Traditional data visualisation performed on local or remote desktop viewers will be also severely challenged in presence of very large data, demanding more efficient rendering strategies, possibly decoupling visualisation and computation, for example moving the latter to a distributed computing infrastructure.

The analysis capabilities offered by existing image viewers are currently limited to the computation of image/region statistical estimators or histograms, and to data retrieval (images or source catalogues) from survey archives. Advanced source analysis, from extraction to catalog cross-matching and object classification, are unfortunately not supported as the graphical applications are not interfaced with source finder batch applications. On the other hand, source finding often requires visual inspection of the extracted catalog, for example to select particular sources, reject false detections or identify the object classes. Integration of source analysis capabilities into data visualisation tools could therefore significantly improve and speed-up the cataloguing process of large surveys, boosting astronomers' productivity and shortening publication times. 

As we approach the SKA era, two main challenges are to be faced in the data visualisation domain: scalability and data knowledge extraction and presentation to users. The present capability of visualisation software to interactively manipulate input datasets will not be sufficient to handle the image data cubes expected in SKA (~200-300 TB at full spectral resolution). This expected volume of data will require innovative visualisation techniques and a change in the underlying software architecture models to decouple the computation part from the visualisation. This is, for example, the approach followed by new-generation viewers such as CARTA \cite{angus_comrie_2019_3403491}, which uses a “tiled rendering” method in a client-server model. In CARTA, storage and computation are carried out on high-performance remote clusters, whereas visualisation of processed products takes place on the client side exploiting modern web features, such as GPU-accelerated rendering. 

However, the expected volume and complexity of SKA data will demand not only enhanced visualisation capabilities but also, principally, efficient extraction of meaningful knowledge, allowing the discovery of new unexpected results. The ability to extract scientific value from large amounts of data indeed represents the SKA ultimate challenge. To address such needs under a unified framework, visual analytics (VA) has recently emerged as the “science of analytical reasoning facilitated by interactive visual interfaces” \cite{yi2007toward}. VA aims to develop techniques and tools to support researchers in synthesising information and deriving insights from massive, dynamic, unclear, and often conflicting data \cite{keim2008visual,ham2010state}. To achieve this goal, VA integrates methodologies from information, geospatial and scientific analytics, and also takes advantage from techniques developed in the fields of data management, knowledge representation and discovery, and statistical analytics. In this context, new developments have been recently done for astronomy. As an example, the encube framework \cite{vohl2016collaborative} was developed to enable astronomers to interactively visualise, compare and query subsets of spectral cubes from survey data. encube provides a large scale comparative visual analytics framework tailored for use with large tiled displays and advanced immersive environments like the CAVE2 \cite{febretti2013cave2} (a modern hybrid 2D and 3D virtual reality environment).

\section{VisIVO Visual Analytics}

VisIVO Visual Analytics \cite{vitello2018vialactea} is an integrated suite of tools focused on handling  massive and heterogeneous volumes of data coming from cutting-edge Milky Way surveys that span the entire Galactic Plane, homogeneously sampling its emission over the whole electromagnetic spectrum. The tool access data previously processed by data mining algorithms and advanced analysis techniques, providing highly interactive visual interfaces that offer scientists the opportunity for in-depth understanding of massive, noisy, and high-dimensional data.

Alongside data collections, the tool exposes also the scientific knowledge derived from the data, including information related to filamentary structures, bubbles and compact sources.

\begin{figure}[ht]
    \centering
    \includegraphics[width=\textwidth]{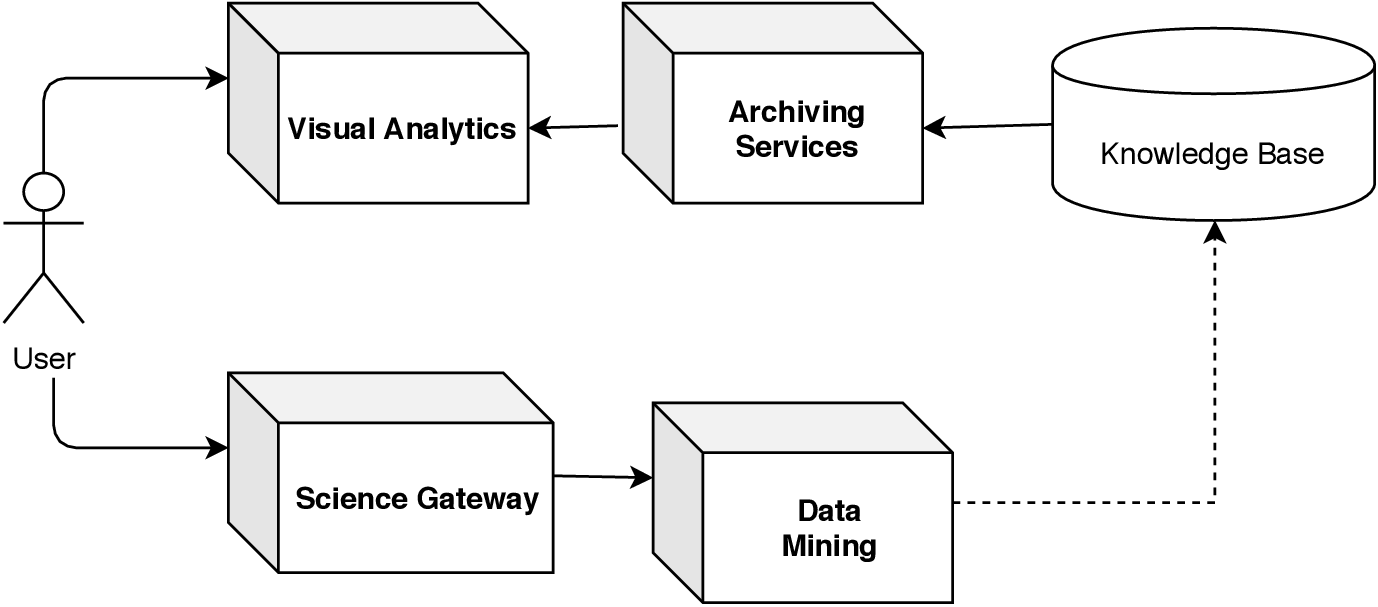}
    \caption{Architecture of the VisIVO Visual Analytics.}
    \label{fig:vlva}
\end{figure}

Figure \ref{fig:vlva} shows the VisIVO Visual Analytics integrated framework, where the Visual Analytics desktop client, the Science Gateway embedding the Data Mining pipelines, and the Knowledge Base can be employed both as independent actors or as interacting components.

\section{EOSCPilot Science Demonstrator}

The EOSCpilot project\footnote{EOSCpilot web page: \url{https://eoscpilot.eu/}} supported the first phase in the development of the European Open Science Cloud, bringing together stakeholders from research infrastructures and e-Infrastructure providers, and engaging with funders and policy makers to propose and trial EOSC's governance framework. 

The VisIVO project has been selected as a science demonstrator \cite{becciani2018eosc} functioning as a high-profile pilot that integrates astrophysical data and visual analytics services and infrastructures, showing interoperability within other scientific domains such as Earth sciences and life sciences. Therefore, the connection with the European Open Science Cloud has been thoroughly investigated, exploiting several services developed within the European Grid Initiative (EGI), such as federated authentication and authorization and the federated cloud for analysis and archiving services. 

The visual analytics application has been further extended by exploiting EOSC technologies for the archive services, as well as intensive analysis employing the ViaLactea Science Gateway\footnote{ViaLactea Science Gateway: \url{https://vialactea-sg.oact.inaf.it/}} \cite{sciacca2017vialactea}.

\begin{figure}[ht]
    \centering
    \includegraphics[width=\textwidth]{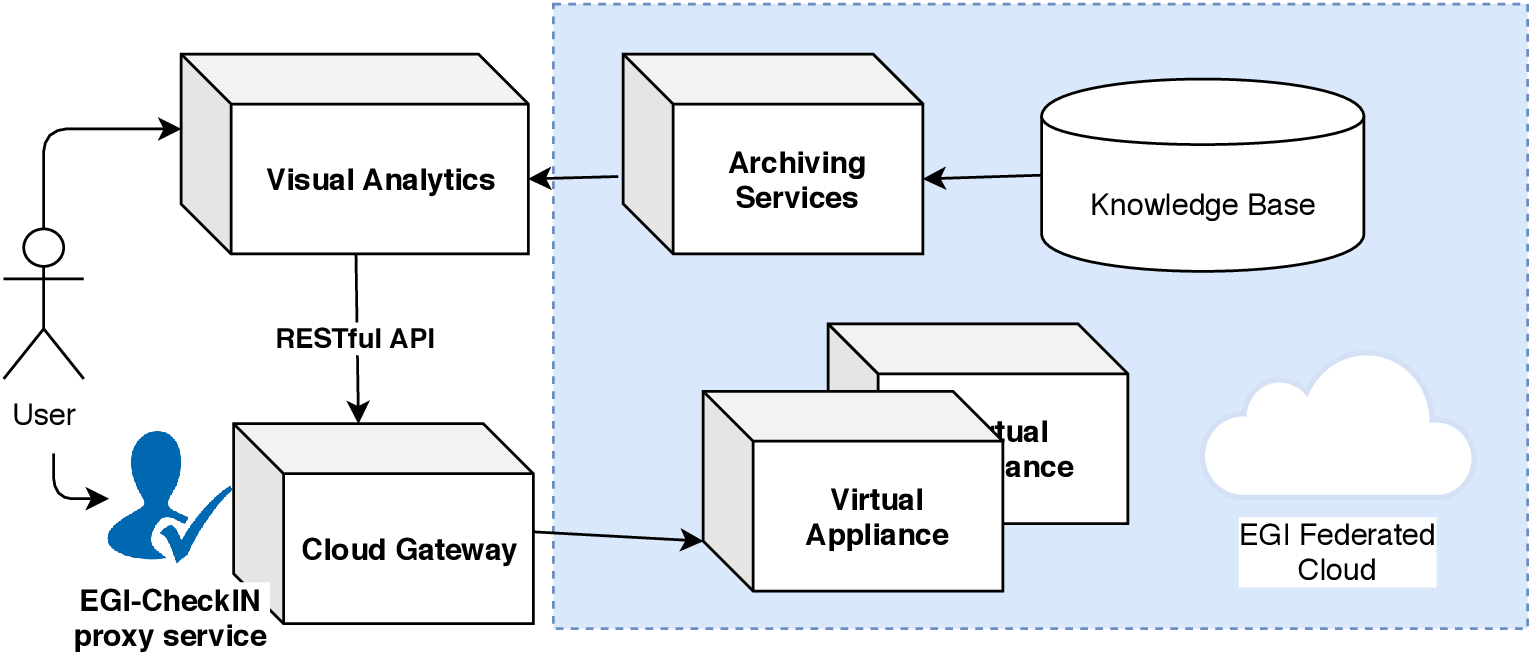}
    \caption{Architecture of the VisIVO EOSC Science Demonstrator implementation and employed services.}
    \label{fig:architecture}
\end{figure}

Figure \ref{fig:architecture} shows the overall architecture of the VisIVO EOSC Science Demonstrator implementation and employed services. The \textit{Archiving Services} (including the knowledge base) have been deployed within the EGI Federated Cloud toward the assurance of a FAIR access to surveys data and related metadata. The \textit{Cloud Gateway} has been integrated with the EGI Check-in\footnote{EGI Check-in service: \url{https://www.egi.eu/services/check-in/}} proxy service to enable the connection from the federated Identity Providers and with the EGI Federated Cloud\footnote{EGI Federated Cloud: \url{https://www.egi.eu/services/cloud-compute/}} to expand the computing capabilities making use of a dedicated \textit{Virtual Appliance} stored into the EGI Applications Database\footnote{EGI Applications Database: \url{https://appdb.egi.eu/store/vappliance/visivo.sd.va}}. The virtual appliance was exploited for massive calculation of spectral energy distributions but may be expanded for more advanced types of analysis in the future.

Furthermore, we have also implemented a lightweight version of the science gateway framework, developing an ad-hoc \textit{RESTful API}, named Cloud for Astrophysics GatEways (CAGE) and available on GitHub\footnote{CAGE codebase: \url{https://github.com/acaland/simple-cloud-gateway}}, to expose a simple set of functionalities to define pipelines and executing scientific workflows on any Cloud resources, hiding all the underlying infrastructures.

\section{Future Works: further EOSC Exploitation}

The H2020 NEANIAS project\footnote{NEANIAS web page: \url{https://www.neanias.eu/}} has been recently approved by the European Commission to address the ‘Prototyping New Innovative Services’ challenge set out in the recent ‘Roadmap for EOSC’ foreseen actions. NEANIAS will drive the co-design, delivery, and integration into EOSC of innovative thematic services, derived from state-of-the-art research assets and practices in three major sectors: underwater research, atmospheric research and space research. Each thematic service will not only address its community-specific needs, but will also enable the transition of the respective community to the EOSC concept and Open Science principles. From a technological perspective, NEANIAS will deliver a rich set of services that are designed to be flexible and extensible; they will be able to accommodate the needs of communities beyond their original definition and to adapt to neighbouring cases, fostering reproducibility and re-usability. 

The foreseen services related to the astrophysics visual analytics are:
\begin{itemize}
\item The \textit{FAIR Data Management and visualisation service} will deliver an advanced operational solution for data management and visualisation for space FAIR data. It will provide tools that enable an efficient and scalable visual discovery, exposed through advanced interaction paradigms exploiting virtual reality. 
\item The \textit{Map Making and Mosaicing of Multidimensional Space Images service} will deliver a user-friendly cloud-based version of the already existing workflow for map making and mosaicing of multidimensional map images based on open source software such as Unimap\cite{piazzo2014unimap} and Montage\cite{berriman2004montage}. It will create multidimensional space maps through novel mosaicing techniques to a variety of prospective users/customers (e.g., mining and robotic engineers, mobile telecommunications companies, space scientists).
\item The \textit{Structure Detection on Large Scale Maps with Machine Learning service} will deliver a user-friendly cloud-based solution for innovative structure detection (e.g. compact/extended sources, filaments), extending the CAESAR\cite{riggi2019c} and CuTEx\cite{molinari2011source} tools with state-of-the-art machine learning frameworks and techniques. The delivered structure detection capabilities will leverage the targeted-users’ opportunities for efficiently identifying and classifying specific structures of interest. 
\end{itemize}

\begin{figure}[ht]
    \centering
    \includegraphics[width=\textwidth]{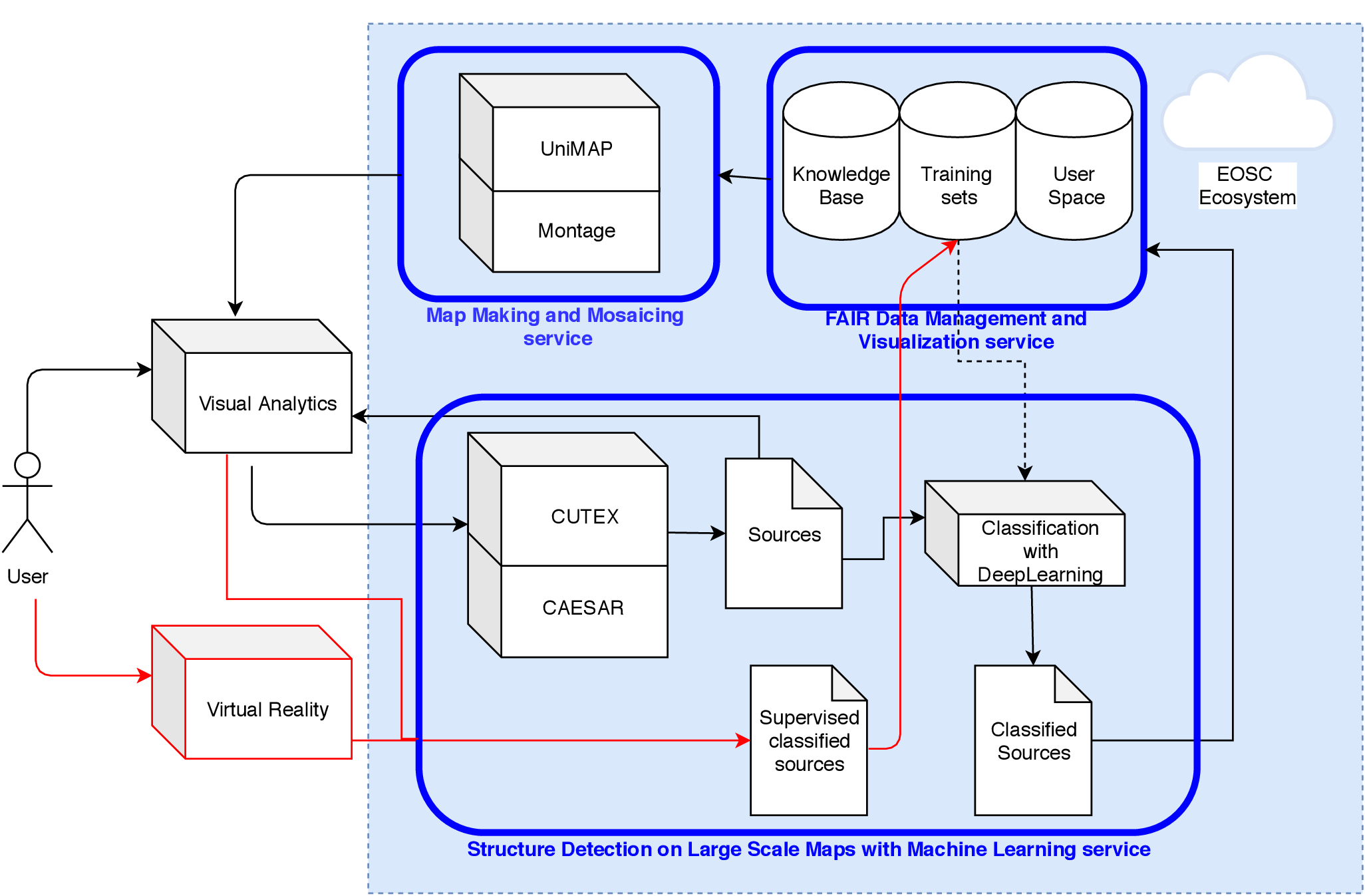}
    \caption{Foreseen NEANIAS Architecture of the visual analytics services in EOSC.}
    \label{fig:architectureNEANIAS}
\end{figure}

The Figure \ref{fig:architectureNEANIAS} shows the main workflow foreseen to exploit the EOSC ecosystem for visualisation, source finding and classification of Big Data images and 3D spectral data cubes coming from multiwavelength Galactic Plane surveys.

The user will employ the Visual Analytics tool to import data from the data management services, opportunely mapped and mosaiced. The tool will exploit the source finding applications to extract sources from the data. Optionally, expert users may employ the Visual Analytics tool and/or Virtual Reality application to interactively classify the sources by visual inspection. The results of this supervised classification can be stored to the data services, enriching the training set for the Deep Learning networks. The extracted sources are then automatically classified with Deep Learning algorithms and the results are stored within the data user space, and optionally published for re-use by other users and/or to enrich the training set.

\section{Conclusion}

We presented the ongoing activities related to the implementation of services, integrated into EOSC, towards addressing the diverse astrophysics user communities needs for visual analytics. The preliminary demonstration implementation developed within the H2020 EOSCPilot project has been summarized. Forthcoming activities to be developed within the H2020 NEANIAS project have been presented, projecting tailored services for FAIR data management and visualisation, multidimensional map creation and mosaicing, and machine learning-supported automated source detection in multidimensional maps.

\section*{Acknowledgements} 

The research leading to these results has received funding from the European Commissions Horizon 2020 research and innovation programme under the grant agreement No. 863448 (NEANIAS).

%
%
\bibliographystyle{splncs03_unsrt}
\bibliography{biblio}  
\end{document}